\documentclass[twocolumn,showpacs,preprintnumbers,amsmath,amssymb]{revtex4}

\usepackage{graphicx}
\usepackage{dcolumn}
\usepackage{bm}
\usepackage{amsmath}
\usepackage{url}

\begin{document}

\title{Neutrino scattering and flavor transformation in supernovae}

\author{John F. Cherry$^{1}$$^{3}$}
\author{J. Carlson$^{2}$$^{3}$}
\author{Alexander Friedland$^{2}$$^{3}$}
\author{George M. Fuller$^{1}$$^{3}$}
\author{Alexey Vlasenko$^{1}$$^{3}$}

\affiliation{$^{1}$Department of Physics, University of California, San Diego, La Jolla, California 92093, USA}
\affiliation{$^{2}$Theoretical Division, Los Alamos National Laboratory, Los Alamos, New Mexico 87545, USA}
\affiliation{$^{3}$Neutrino Engineering Institute, New Mexico Consortium, Los Alamos, New Mexico 87545, USA}

\date{March 6, 2012}

\begin{abstract}
We argue that the small fraction of neutrinos that undergo direction-changing scattering outside of the neutrinosphere could have significant influence on neutrino flavor transformation in core-collapse supernova environments.  We show that the standard treatment for collective neutrino flavor transformation is adequate at late times, but could be inadequate in the crucial shock revival/explosion epoch of core-collapse supernovae, where the potentials that govern neutrino flavor evolution are affected by the scattered neutrinos.  Taking account of this effect, and the way it couples to entropy and composition, will require a new paradigm in supernova modeling.
\end{abstract}
\preprint{LA-UR-12-10419}

\maketitle


In this letter we point out a surprising feature of neutrino flavor transformation in core-collapse supernovae.  These supernovae have massive star progenitors which form cores which collapse to nuclear density and produce proto-neutron stars.  The gravitational binding energy released, eventually some $\sim 10\,\%$ of the rest mass of the neutron star, is emitted as neutrinos of all flavors in a time window of a few seconds.  
Diverting a small fraction of this neutrino energy into heating can drive revival of the stalled core bounce shock~\cite{Arnett:1977kx,Bowers:1982vn,Bethe:1985rt,Blondin:2003ve,Blondin:2007qy,Scheck:2008ul,Brandt:2011lr} creating a supernova explosion and setting the conditions for the synthesis of heavy elements~\cite{Blondin:2003ve,Arcones:2007fk,Scheck:2008ul,Hammer:2010yq,Brandt:2011lr}.  However, the way neutrinos interact in this environment depends on their flavors, necessitating calculations of neutrino flavor transformation.  These calculations show that neutrino flavor transformation has a rich phenomenology, including collective oscillations~\cite{Fuller87,Fuller:1992eu, Balantekin:2000hl, Pastor:2002zl,Balantekin05, Duan06a, Duan06b, Hannestad:2006qd,Fuller06,Duan06c,Duan07a,Duan07b,Duan07c,friedland:2006lr,Esteban07,Balantekin:2007kx,Fogli:2007ys, Duan08,Duan:2008eb,Duan:2008qy,Dasgupta09,Cherry:2010lr,Cherry:2011bh,Duan:2010fr,Duan:2011fk,Pehlivan:2011rt, Banerjee:2011lr,Dasgupta:2011uq,Duan:2012qd}, which can affect important aspects of supernova physics
~\cite{Gil-Botella03,Duan06a,Duan06b,Duan06c,Duan07a,Duan07b,Duan07c,friedland:2006lr,Lunardini08,Duan08,Duan:2008eb,Duan:2008qy,Kneller:2008rt,Gava:2009yq,Cherry:2010lr,Cherry:2011bh,Kneller:2010ys}.  For example, neutrino-heated heavy element r-process nucleosynthesis~\cite{Qian95,Balantekin:2006yq,Qian:2007lr,Ning07,Duan:2011lr} and potentially supernova energy transport above the core and the explosion itself~\cite{Fuller:1992eu,Murphy:2008qf,Dasgupta:2011uq} could be affected.  

\begin{figure}[h]
\centering
\includegraphics[scale=.32]{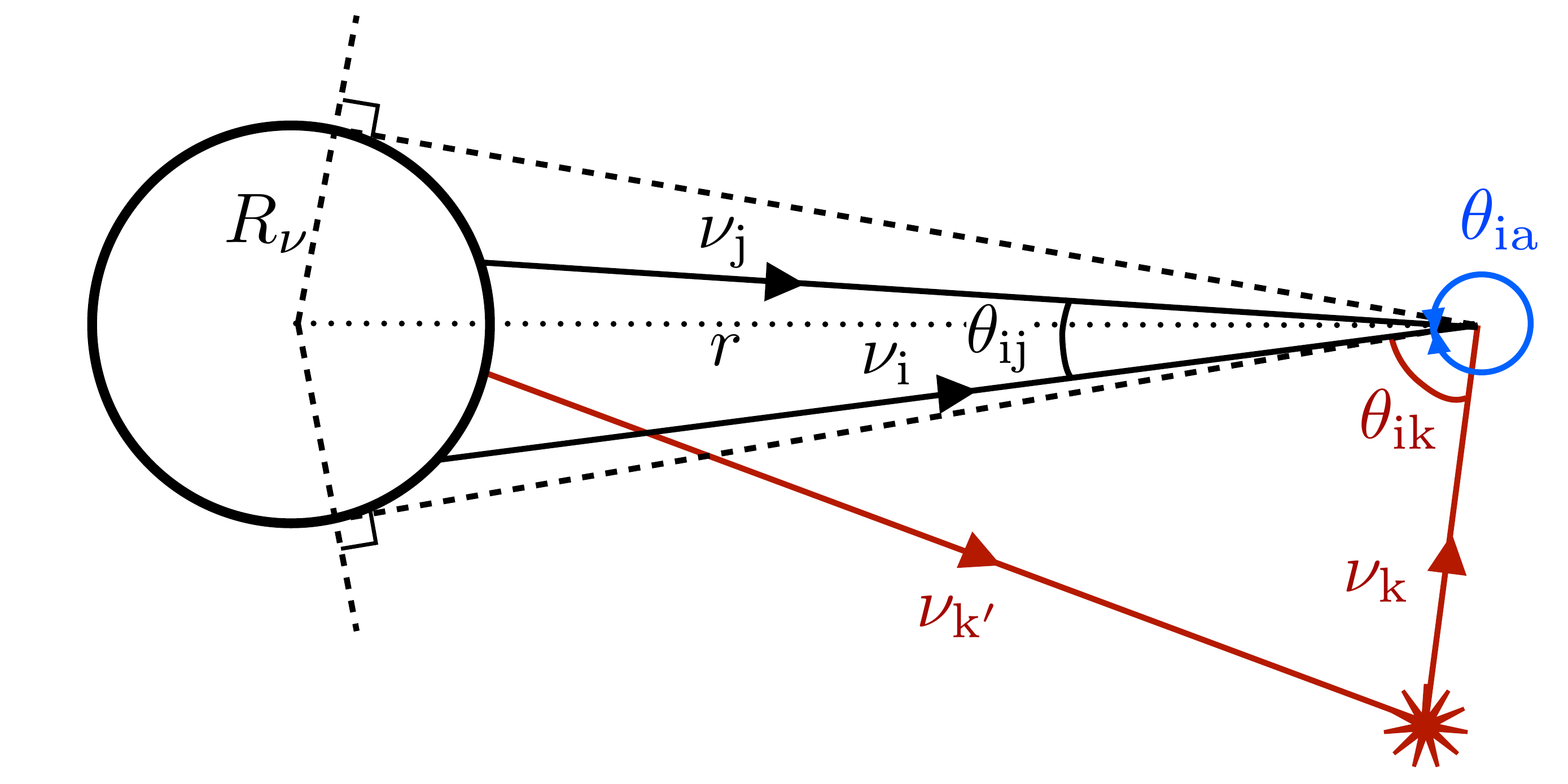}
\caption{Supernova neutrino emission geometry.}
\label{fig:Cartoon}
\end{figure}

All collective neutrino flavor transformation calculations employ the \lq\lq Neutrino Bulb\rq\rq\ model, where neutrino emission is sourced from a \lq\lq neutrinosphere\rq\rq , taken to be a hard spherical shell from which neutrinos freely stream.  This seems like a reasonable approximation because well above the neutrinosphere scattered neutrinos comprise only a relatively small fraction of the overall neutrino number density.  However, this optically thin \lq\lq halo\rq\rq\ of scattered neutrinos nonetheless may influence the way flavor transformation proceeds.  This result stems from a combination of the geometry of supernova neutrino emission, as depicted in Fig.~\ref{fig:Cartoon}, and the neutrino intersection angle dependence of neutrino-neutrino coupling.

Neutrinos are emitted in all directions from a neutrinosphere of radius $R_\nu$, but those that arrive at a location at radius $r$, and suffer only forward scattering, will be confined to a narrow cone of directions (dashed lines in Fig.~\ref{fig:Cartoon}) when $r\gg R_\nu$.  In contrast, a neutrino which suffers one or more direction-changing scattering events could arrive at the same location via a trajectory that lies well outside this cone.

Following neutrino flavor evolution in the presence of scattering, in general, requires a solution of the quantum kinetic equations~\cite{Sigl93, Strack:2005rr,Vlasenko:2012lr}.  However, the rare nature of the scattering that generates the halo suggests a separation between the scattering-induced and coherent aspects of neutrino flavor evolution.  In the coherent limit the neutrino-neutrino Hamiltonian, $\hat{H}_{\nu\nu}$, couples the flavor histories for neutrinos on intersecting trajectories~\cite{Sigl93,Qian95,Pastor02,Duan:2010fr}.  As shown in Fig.~\ref{fig:Cartoon}, a neutrino $\nu_{\rm i}$ leaving the neutrinosphere will experience a potential given by a sum over neutrinos and antineutrinos located at the same point as neutrino $\nu_{\rm i}$:
\begin{eqnarray}
& \hat{H}_{\nu\nu} & 
 =  \sqrt{2}\,G_{\rm F}\,\sum_{\rm a}{\left( 1-\cos{\theta_{\rm ia}}\right) n_{\nu,\rm a}\, \vert\psi_{\nu,\rm a}\rangle\, \langle\psi_{\nu,\rm a}\vert} 
\nonumber
\\
                 &  - &\sqrt{2}\,G_{\rm F}\, \sum_{\rm a}{\left( 1-\cos{\theta_{\rm ia}} \right) n_{\bar\nu,\rm a}\, \vert\psi_{\bar\nu,\rm a}\rangle\, \langle\psi_{\bar\nu,\rm a}\vert},
 \label{Hnunu}
\end{eqnarray}
where the flavor state of neutrino $\nu_{\rm a}$ is represented by $\vert\psi_{\nu,\rm a}\rangle$, and $\theta_{\rm ia}$ is the angle of intersection between $\nu_{\rm i}$ and neutrino or antineutrino $\nu_{\rm a}/\bar{\nu}_{\rm a}$.  Here $n_{\nu,\rm a}$ is the local number density of neutrinos in state $\rm a$, and the $1-\cos\theta_{\rm{ia}}$ factor disfavors small intersection angles, thereby suppressing the potential contribution of the forward-scattered-only neutrinos~\cite{Fuller87,Fuller:1992eu}.  Direction-altered scattered neutrinos may have larger intersection angles as shown in Fig.~\ref{fig:Cartoon}, and therefore can contribute significantly to the flavor-changing potentials, despite their small numbers.

In the mean-field, coherent approximation, neutrino flavor evolution is governed by a Schr\"odinger-like equation~\cite{Halprin:1986cr}, $i {{\partial \vert\psi_{\nu,\rm i}\rangle  }/{\partial t  }}=\hat{H} \vert\psi_{\nu,\rm i}\rangle$, where  $t$ is an Affine parameter along neutrino $\nu_{\rm i}$'s world line, and $\hat{H} = \hat{H}_{\rm V} + \hat{H}_{\rm e} + \hat{H}_{\nu\nu}$ is the appropriate neutrino propagation Hamiltonian, with vacuum and  matter components $\hat{H}_{\rm V}$ and $\hat{H}_{\rm e} $, respectively.  $\hat{H}_{\nu\nu}$ can be split into two pieces: $\hat{H}_{\nu\nu}^{\rm bulb}$, contributed by neutrinos (index $\rm j$ in Fig.~\ref{fig:Cartoon}) which propagate directly (straight lines) from the surface of the neutrinosphere; and $\hat{H}_{\nu\nu}^{\rm halo}$, contributed by neutrinos that suffer direction-changing scattering outside the neutrinosphere (index $\rm k$ in Fig.~\ref{fig:Cartoon}) and propagate coherently thereafter.  To wit, $\hat{H}_{\nu\nu} = \hat{H}_{\nu\nu}^{\rm bulb} + \hat{H}_{\nu\nu}^{\rm halo}$.

The operators $\hat{H}_{\nu\nu}^{\rm halo}$ and $\hat{H}^{\rm bulb}_{\nu\nu}$ depend on the complex phases of the neutrino flavor states which contribute to them, so that the relative leverage of these operators in determining flavor transformation at any point requires numerical calculations.  Some conditions have been shown to give phase locking, while other conditions give phase decoherence~\cite{Qian95,Duan06b,Esteban07,Duan:2010fr,Duan:2012qd}.  For the purpose of evaluating the validity of the Neutrino Bulb model, we ignore path length difference-induced phase averaging~\cite{Qian95} and compute the  {\it maximum} magnitude of the diagonal Hamiltonian elements, which we denote with $\vert \hat{H}_{\nu\nu}^{\rm halo}\vert$ and $\vert \hat{H}^{\rm bulb}_{\nu\nu}\vert$.  A necessary condition for the validity of the Neutrino Bulb model is that $\vert \hat{H}_{\nu\nu}^{\rm bulb}\vert \gg \vert  \hat{H}_{\nu\nu}^{\rm halo}\vert$. 

A simple argument can be made about which varieties of spherically symmetric density profiles could render the Neutrino Bulb model inadequate.  Consider a series of spherical shells of matter stacked around the neutrinosphere.  These shells are taken to isotropically scatter neutrinos, and, as we discuss below, neutral current neutrino-nucleon/nucleus scattering does just this.  Some of these neutrinos will contribute number density and flavor information to the sum in Eq.~\ref{Hnunu}, adding to $\hat{H}_{\nu\nu}^{\rm halo}$.  For a point $r$ well outside of these shells, the number density of neutrinos being scattered to this location from a shell at radius $r^\prime$, multiplied by the average value of $\left( 1 - \cos\theta_{\rm ik}\right)$ for neutrinos coming from this shell, is $\propto \rho\left( r^\prime\right)\delta r^\prime \left(r^\prime / R_{\nu}\right)^2$, where $\delta r^\prime$ is the thickness of the shell.  As $r^\prime$ approaches $r$, the contribution from these shells is regulated.  For small $r^\prime$, the shell contributions are regulated by the neutrinosphere.

These considerations imply that when $\rho\left( r^\prime\right)\propto {r^\prime}^{-3}$, the potential contributed by a given shell will be $\propto \delta r^\prime / r^\prime \propto \delta \log \left( r^\prime\right)$.  Any selection of logarithmically spaced shells with $r^\prime < r$ will cause each shell to make an equal contribution of neutrino number density at $r$.  Physically, we might expect density features of size $r^\prime$ at radius $r^\prime$.  This matter density configuration will cause the ratio of $\vert \hat{H}_{\nu\nu}^{\rm halo}\vert /\vert \hat{H}^{\rm bulb}_{\nu\nu}\vert$ to remain fixed with increasing radius.   

\begin{figure}[h]
\centering
\includegraphics[scale=.65]{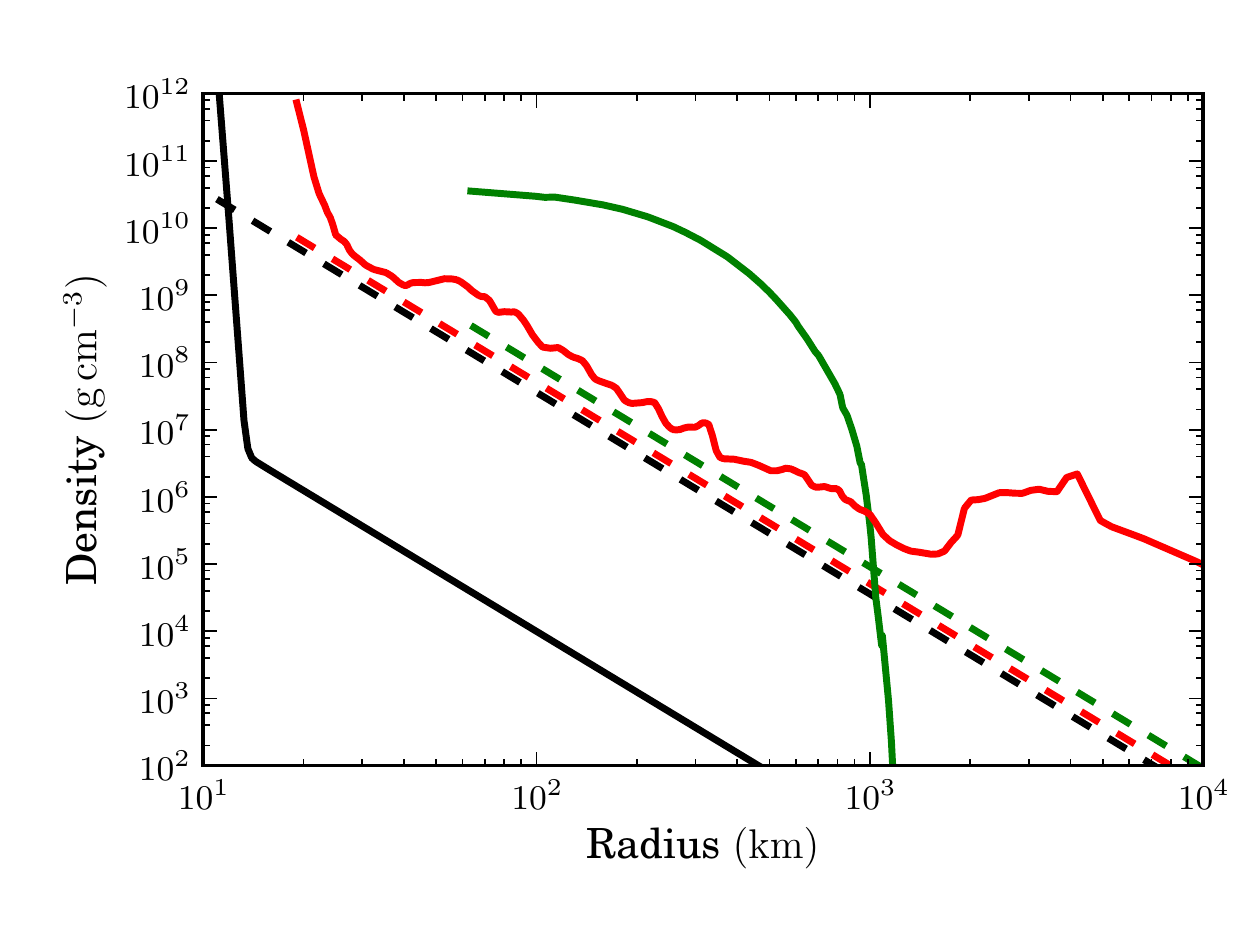}
\caption{Solid lines show matter density profiles and dashed lines the corresponding Neutrino Bulb ($1\,\%$) safety criteria from Eq.~\ref{Safety}.  Black lines are for the late-time neutrino driven wind environment~\cite{Duan06a}, green lines the neutronization burst O-Ne-Mg core-collapse environment~\cite{Nomoto84,Nomoto87}, and red lines the Fe-core-collapse shock revival environment~\cite{Bruenn:2009uq}.}
\label{fig:1D}
\end{figure}

\begin{figure*}[ht]
\centering
\includegraphics[scale=0.37]{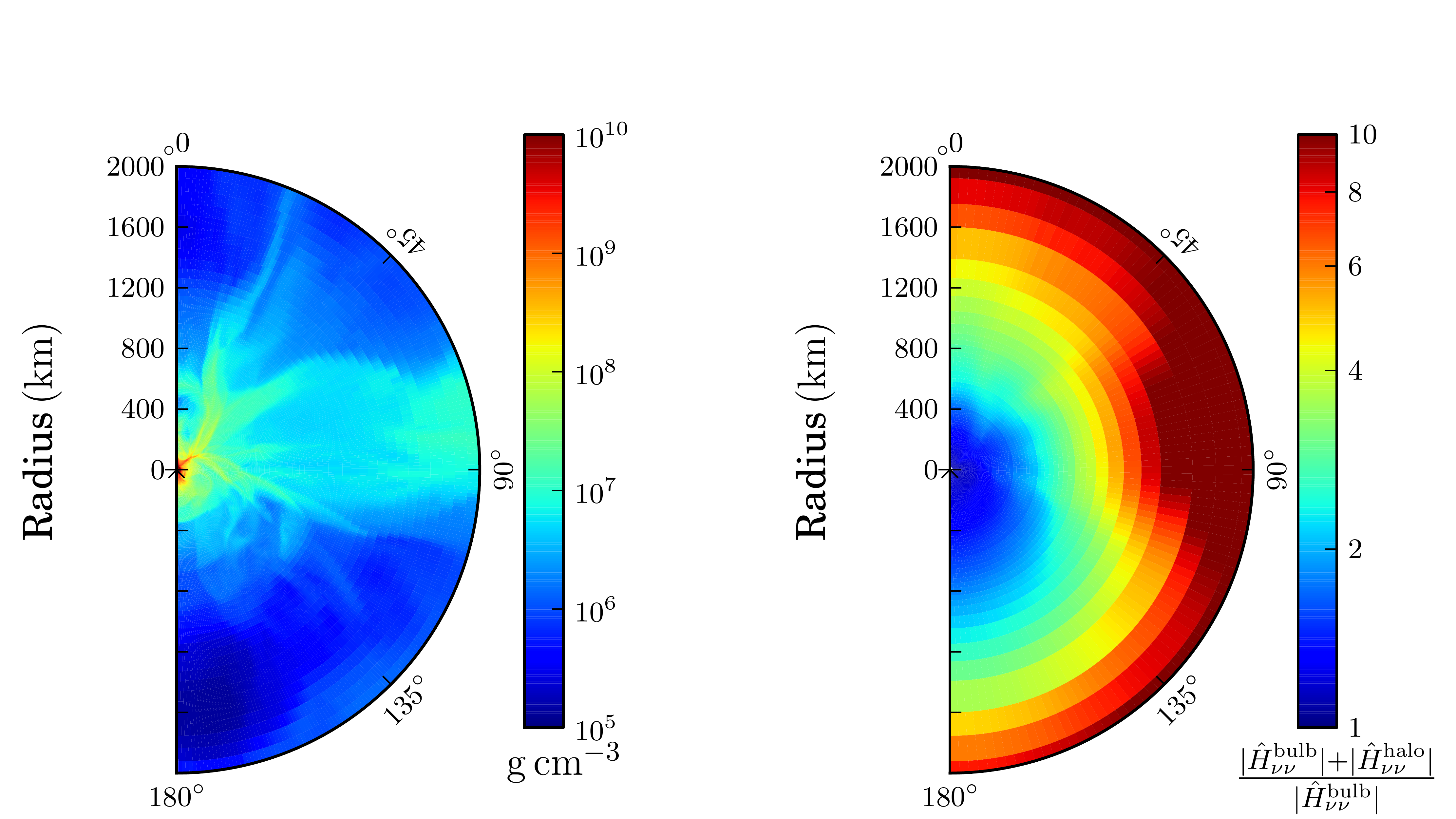}
\caption{Left: Color scale indicates the density within the shock front in a $15\, M_\odot$ progenitor core-collapse supernova $500\,\rm ms$ after core bounce, during the shock revival epoch~\cite{Bruenn:2009uq}.  Right:  Effect of the scattered neutrino halo for the matter distribution at Left.  Color scale indicates the ratio of the sum of the maximum (no phase averaging) magnitudes of the constituents of the neutrino-neutrino Hamiltonian, $\vert\hat{H}_{\nu\nu}^{\rm bulb}\vert +\vert\hat{H}_{\nu\nu}^{\rm halo}\vert$, to the contribution from the neutrinosphere $\vert\hat{H}_{\nu\nu}^{\rm bulb}\vert$.}
\label{fig:2DHvv}
\end{figure*}

To compare the contributions of the halo shells to $\vert \hat{H}_{\nu\nu}^{{\rm bulb}}\vert$, we observe that the neutrinosphere (more precisely, the \emph{transport sphere}~\cite{Raffelt:2001ul}) can be treated in the same spirit.  The transport sphere is characterized by the neutrino optical depth, $\tau$, equal to unity. 
Requiring that the logarithmic shells above the neutrinosphere contribute much less than the neutrinosphere itself results in
\begin{equation}
\rho\left( r\right) \ll {\rho_{\tau=1}}\left(\frac{R_\nu}{r}\right)^3\ .
\label{Safety}
\end{equation}
Early in the explosion epoch, the transport sphere corresponds to physical radii $R_{\nu}\sim 30-60$ km and densities $\rho_{\tau=1} \sim 10^{\left(11\ {\rm to}\ 12\right)}\, {\rm g\, cm^{-3}}$~\cite{Raffelt:2001ul}.  In fact, Eq.~\ref{Safety} assumes that the thickness $\Delta R_{\nu}$ of the neutrinosphere is $\sim R_{\nu}$, whereas models show that $\Delta R_\nu < R_\nu$, implying a more stringent constraint by a factor of $\Delta R_{\nu}/R_\nu$.

If $\vert \hat{H}_{\nu\nu}^{\rm halo}\vert /\vert \hat{H}^{\rm bulb}_{\nu\nu}\vert < 1\,\%$ is taken as the limit where $\hat{H}_{\nu\nu}^{\rm halo}$ can be neglected, then by Eq.~\ref{Safety} the range of density profiles for which the Neutrino Bulb model is likely to be adequate is $\rho\left( r\right) < 0.01\times {\rho_{\tau=1}}\left(R_\nu /r \right)^3$.  As long as the matter density in the supernova remains below this limit, there is no danger that the fractional potential contribution from the scattered halo, $\vert\hat{H}_{\nu\nu}^{\rm halo}\vert$, will grow above $1\,\%$.  Fig.~\ref{fig:1D} shows the density profiles for several core-collapse supernova environments alongside the corresponding $1\,\%$ safety criterion for each profile.

As is evident in Fig.~\ref{fig:1D}, the Fe-core-collapse shock revival environment will have a significant scattered halo.  Even though the O-Ne-Mg core-collapse density profile~\cite{Nomoto84,Nomoto87} drops into the safe zone at $r > 1000\,\rm{km}$, these models nevertheless will have a significant scattered halo originating from shells at lower radius where the density curve is above the $1\,\%$ safety margin.  Only late-time neutrino-driven wind models avoid scattered halo complication~\cite{Duan06a,Kitaura:2006zr,Fogli:2007ys,Dasgupta09,Friedland:2010gf,Fischer:2010lq,Duan:2011fk}.  Fe-core-collapse models (e.g., the red curve in Fig.~\ref{fig:1D}), in general, exhibit an average density profile that is $\propto r^{-\left(2\ \rm to\ 3\right)}$, which means that $\vert\hat H^{\rm halo}_{\nu\nu}\vert /\vert\hat{H}^{\rm bulb}_{\nu\nu}\vert$ is expected to increase with radius.  Note, however, that though the relative contribution of the halo may grow with radius, at sufficiently large distance from the proto-neutron star the neutrino-neutrino potential ceases to be physically important.

Matter inhomogeneity, an essential feature of supernova explosion models~\cite{Herant:1994qf,Burrows:1995pd, Scheck:2008ul,Blondin:2003ve,Blondin:2007qy, Bruenn:2009uq, Brandt:2011lr}, adds complexity to this issue.  To study this effect we use the 2D matter density distribution, Fig.~\ref{fig:2DHvv}, taken from a supernova model derived from a $15\, M_{\odot}$ progenitor~\cite{Bruenn:2009uq}.  This snapshot corresponds to $500\,\rm{ms}$ after core bounce, during the shock revival epoch, after the onset of the SASI~\cite{Blondin:2003ve,Blondin:2007qy}.  We mock up a full 3D density profile by cloning the 2D profile into a 3D data cube.  Starting with an initial flux of neutrinos from the neutrinosphere~\cite{Keil:2003qy}, and taking all baryons to be free nucleons, we use the full energy dependent neutral current neutrino-nucleon scattering cross sections~\cite{Tubbs:1975ve} to calculate the number flux of neutrinos scattered out of each spatial zone and into every other spatial zone (retaining the necessary information about relative neutrino trajectories between zones).  We compute the magnitude of $\vert\hat{H}_{\nu\nu}^{\rm halo}\vert$ at each location in the 2D slice that comprises the original density distribution.

In this example calculation the scattered halo is taken to be composed of neutrinos which have suffered only a single direction-changing scattering.  Because the halo region is optically thin for neutrinos, multiple scatterings become increasingly rare with radius and do not have a geometric advantage in their contribution to $\vert \hat{H}_{\nu\nu}^{\rm halo}\vert$ relative to singly-scattered neutrinos.  Neutrinos which experience direction-changing scattering that takes them into the same cone of directions as neutrinos forward scattering from the neutrinosphere are counted as contributing to the halo (these neutrinos contribute $\sim 10^{-6}$ of the halo potential).  As before, we neglect the effects of neutrino flavor oscillations.  Fig.~\ref{fig:2DHvv} shows the results of this calculation out to a radius of $r = 2000\, \rm km$.  Disturbingly, neutrinos from the scattered halo in this 2D model nowhere contribute a maximum magnitude less than $14\,\%$ of the neutrino-neutrino potential magnitude, and in many places contribute $90\,\%$ or more of the total.  Fig.~\ref{fig:2DHvv} shows that matter inhomogeneities generate large corresponding scattered halo inhomogeneities. 

The inhomogeneity of the scattered halo is increased by several scattering processes which have been omitted from this illustrative calculation.  We did not include neutrino-electron scattering.  This scattering process has smaller cross sections and relatively forward peaked angular distributions and therefore produces a subdominant contribution to $\vert\hat{H}_{\nu\nu}^{\rm halo}\vert$.  What is more important is that our calculation leaves out what is likely the dominant source of neutrino direction-changing scattering in the low entropy regions of the supernova envelope: coherent neutrino-nucleus neutral current scattering.  

The cross sections for this process scale as the square of the neutrino energy and square of the nuclear mass number $A$.  In fact, since the proper number density of nuclear targets is $\propto A^{-1}$, but the coherent scattering cross section $\propto A^2$, the overall scattered halo potential contribution stemming from this process is $\propto A$.  This process, like neutral current neutrino-nucleon scattering, is flavor independent and flavor preserving, simply changing neutrino direction.  

Since the heavy nucleus mass fraction, and the distribution of nuclear mass numbers, can depend sensitively on the entropy and electron fraction~\cite{Bethe:1979mz}, coherent neutral current scattering could couple neutrino flavor transformation to macroscopic, multi-dimensional structures in the supernova envelope.  For example, the model shown in Fig.~\ref{fig:2DHvv} has relatively lower entropy, downward-flowing, higher nuclear mass fraction matter; and higher entropy, upward-flowing plumes, with relatively lower nuclear mass fraction.  This could produce a scattered halo with a complicated 3D geometry and flavor content, creating a non-trivial enhancement to the inhomogeneities evident in the scattered halo potential shown in Fig.~\ref{fig:2DHvv}.  

Because scattering processes are energy dependent, neutrinos in the scattered halo possess different energy spectra than forward-scattered only neutrinos.  Furthermore, the flavor content of the scattered halo will not match that of neutrinos emerging from the neutrinosphere.  For example, consider emergent $\nu_{\rm e}$ and $\bar{\nu}_{\rm e}$ number fluxes that are equal, yet have different energy spectra.  Taken alone, unmolested by neutrino flavor oscillations, these fluxes give $\vert\hat H^{\rm bulb}_{\nu\nu}\vert = 0$.  However, passing through the energy-dependent scattering processes, they yield $\vert\hat{H}_{\nu\nu}^{\rm halo}\vert \neq 0$.

Anticipating the course of neutrino flavor evolution in this environment is clearly challenging.  $\vert \hat H_{\rm e}\vert$ is by far the largest contribution to $\vert \hat H \vert$ during the shock revival epoch.  However, considering the Bulb neutrinos alone, current coherent calculations show that neutrino collective flavor oscillations can proceed despite a large matter potential~\cite{Hannestad:2006qd,Duan06c,Duan06a,Duan07a}.  The criterion for the matter suppression of collective oscillations~\cite{Esteban-Pretel:2008bh}, $\Delta\vert \hat H_{\rm e}\vert \sim \Delta\vert \hat H_{\rm V} \vert$, where $\Delta$ denotes the {\it dispersion} in these potentials for Bulb neutrinos, is not met where the matter densities in Fig.~\ref{fig:2DHvv} drop below $\sim 10^{7\ \rm to\ 8}\, \rm g\,\rm cm^{-3}$ (the wide range is due to the geometric dependence of $\Delta\vert \hat H_{\rm e}\vert $).  In a further complication, neutrinos from the spatially extended scattered halo could arrive at a given location along many different trajectories with different path lengths, so that significant neutrino oscillation phase averaging~\cite{Qian95} could come into play.  This has been shown to suppress collective oscillations in some conditions~\cite{Duan:2011fk}.  Inhomogeneity and the intersection angle dependence of the neutrino-neutrino interaction may make phase averaging incomplete.  Ascertaining the role of decoherence and phase averaging processes requires detailed calculation with specific supernova models~\cite{Duan:2012qd}.  Even if collective oscillations are found to be suppressed at small radius, they may be operating, e.g. above the shock, because the halo extends the collective oscillation region.

Though validating coherent flavor transformation studies for late-time neutrino driven wind models, our calculations demonstrate the potential inadequacy of these treatments in an environment important for the understanding of the supernova explosion mechanism and nucleosynthesis.  Ultimately, the scattered halo changes the nature of the neutrino flavor transformation problem: it broadens the region influencing flavor evolution from just the neutrinosphere to a much larger fraction of the supernova envelope; and it introduces essential multi-dimensional effects.  The standard Neutrino Bulb model by its nature is an initial value problem at each radius $r$, while the scattered halo makes it necessary to consider how flavor transformations at large radii can feed back into the evolution at smaller radii.  With this additional source of nonlinearity, qualitatively new phenomena could, in principle, occur.  Further, the extended scattered halo can couple neutrino flavor evolution to the nuclear composition and complex 3D flow geometries which are characteristics of the supernova explosion epoch.  A self-consistent solution of this problem likely will demand new computational capabilities and approaches.  Given the importance of neutrinos and the supernova phenomenon for so many aspects of our understanding of the cosmos, it may be that there is no choice but to seek such a solution.

This work was supported in part by NSF grant PHY-09-70064 at UCSD, and by the DOE Office of Science, the LDRD Program and Open Supercomputing at LANL.  We thank B. Messer and A. Mezzacappa for providing their supernova model. We would like to thank V. Cirigliano, Y.-Z. Qian, the Topical Collaboration for Neutrinos and Nucleosynthesis in Hot and Dense Matter at LANL, and the New Mexico Consortium.

\bibliography{allref}

\end{document}